\input epsf
\documentstyle[floats,eqsecnum,aps]{revtex}
\tighten
\begin{document}
\draft
\twocolumn[\hsize\textwidth\columnwidth\hsize\csname @twocolumnfalse\endcsname
\preprint{AEI-1999-6}

\title{Perturbative evolution of nonlinear initial data for
binary black holes:\\
Zerilli vs. Teukolsky}
\author{Carlos O. Lousto}
\address{Albert-Einstein-Institut,
Max-Planck-Institut f\"ur Gravitationsphysik,\\
Am M\"uhlenberg 5, D-14476 Golm, Germany\\
}
\date{\today}
\maketitle
\begin{abstract}

We consider the problem of evolving nonlinear initial data 
in the close limit regime. Metric and curvature perturbations
of nonrotating black holes are equivalent to first perturbative order,
but Moncrief waveform in the former case and Weyl scalar $\psi_4$
in the later differ when nonlinearities are present.
For exact Misner initial data (two equal mass black holes 
initially at rest), metric perturbations evolved via 
the Zerilli equation suffer of a premature break down 
(at proper separation of the holes $L/M\approx2.2$) while the 
exact Weyl scalar $\psi_4$ evolved via the Teukolsky 
equation keeps a very good agreement with
full numerical results up to $L/M\approx3.5$.
We argue that this inequivalent behavior holds for a wider
class of conformally flat initial data than those studied here.
We then discuss the relevance of these results for second order perturbative
computations and for perturbations to take over full numerical evolutions
of Einstein equations.

\end{abstract}
\vskip2pc]

\section{Introduction}

There is a revival of the interest on perturbation theory of black
holes since the work of Price and Pullin\cite{PP94} who put forward
the close limit approximation. This approach considers the final merger
stage of binary black holes as a single, perturbed black hole.
They studied the Misner problem, two equal mass black holes initially
at rest, and compared their
results with full numerical evolution of Einstein equations for various
initial separations. The impressive agreement, even for not so
small separations, triggered several researchers to test these ideas
for initial data representing boosted towards each other, single
spinning plus Brill waves, and orbiting black holes\cite{P98}.

Abrahams and Price\cite{AP96a} found that if one does not linearize
Misner initial data, but directly extract the $\ell$-multipoles from
the exact 3-metric, the close limit approximation breaks down for much
smaller separations; in a regime where the linearized approach precisely
agrees with full numerical calculations. In this paper we shed some
light on the origin of this surprising result.

Price and Pullin used the Zerilli equation in the time domain
to evolve perturbations around a Schwarzschild black hole.
There is an alternative
approach to deal with perturbations of black holes (even with net rotation)
due to Teukolsky\cite{T73} based on the Newman-Penrose 
formalism. The two approaches are related and equivalent when one deals
with first order perturbations\cite{CL98,CKL98}. Here we want to explore how
they behave under nonlinear components present in the initial data.
This study is relevant to test the idea\cite{LP00} of using perturbation theory
at the final merger stage of binary black holes taking over Cauchy data
from full numerical simulations that started when the two black holes
were fully detached (for instance near the ISCO). Our results will also
serve as precise analytic benchmarks for this case.
It is also useful in order to reliably
use second order perturbation technics when initial data do not come
in an analytic form, but as a numerical table, thus making
a complicated if not an impossible task to disentangling each 
perturbative order.

\section{Initial Data}

Misner\cite{M63} found a solution to the conformally flat, time symmetric
initial value problem representing two black holes at rest separated by
a proper distance $L$ parametrized by $\mu_0$ (see below). For $\mu_0<1.8$
$(L/M<3.3)$ a common event horizon encompasses the system, and for $\mu_0<1.36$
$(L/M<2.5)$ a common apparent horizon appears.

The Misner three-geometry takes the form
\begin{equation}\label{M3}
  ds^2_{\rm Misner} = \Phi(r,\theta)^4
  \left(\frac{dr^2}{1-2M/r}+r^2d\theta^2+r^2\sin^2\theta\ d\varphi^2\right)
\end{equation}
where
\begin{equation}
   {\Phi}=\frac{\phi}{\phi_0}\ ;\quad \phi_0\dot=\left(1+\frac{M}{2R}\right).
\end{equation}

Here we identified $R$, the conformal space radial coordinate,
with the Schwarzschild isotropic coordinate,
\begin{equation}\label{rtrans}
   R =\bar{r}\dot=\frac{1}{4}\left(\sqrt{r}+\sqrt{r-2M}\right)^2.
\end{equation}

The conformal factor $\phi$ is given by,
\begin{eqnarray}\label{genfn}
  \phi &=&1+ \delta\sum_{n\neq0}
\left[(1+\delta^2)\sinh^2{n\mu_0}\right. \\
&&\quad \left.+\delta\cos\theta\sinh{2n\mu_0}
        +\delta^2 \right]^{-1/2},\nonumber
\end{eqnarray}
where $\delta=M/(4R\Sigma_1)$, and
\begin{equation}
       \Sigma_1\dot=\sum_{n=1}^\infty\frac{1}{\sinh{n\mu_0}}.
\end{equation}
This metric represents an asymptotically flat three geometry
with total ADM mass $M$.

The proper distance $L$ between the throats can be written as
\begin{equation}
 L = \frac{M}{2\Sigma_1}
\left(1+2\mu_0\sum_1^\infty \frac{n}{\sinh{n\mu_0}} \right).
\end{equation}

\section{Metric perturbations approach}

The theory of metric perturbations around a Schwarzschild black hole was 
originally derived by Regge and Wheeler for odd-parity 
perturbations and by Zerilli for even-parity ones.  
The spherically symmetric background allows for a multipole 
decomposition, labeled by $(\ell m)$, 
even in the time domain. Moncrief\cite{M74} has 
given a gauge-invariant formulation of the problem
in terms of the three-geometry perturbations. For the head-on collision of
two black holes, that concern us here, only even parity modes are present.
They are described by a single wave (Zerilli's) equation
\begin{equation}\label{rtz}
-\frac{\partial^2Q^{(\ell m)}}{\partial t^2}
+\frac{\partial^2Q^{(\ell m)}}{\partial r*^2}-V_{\ell}(r)Q^{(\ell m)}=0 \; .
\end{equation}
Here $r^*\dot= r+2M\ln(r/2M-1)$, and the potential 
\begin{eqnarray}\label{zpotential}
V_\ell(r)&=&\left(1-\frac{2M}{r}\right)\\
&&\times\frac{2\lambda^2(\lambda+1)r^3+6\lambda^2Mr^2+18\lambda M^2r+18M^3
}{r^3(\lambda r+3M)^2 },\nonumber
\end{eqnarray}
where ${\lambda}\dot=(\ell+2)(\ell-1)/2.$
The Moncrief wave function $Q^{(\ell m)}$,
in terms of the metric perturbations in the Regge--Wheeler notation, is
\begin{eqnarray}\label{psidef}
    Q^{(\ell m)}&=&\frac{r}{{\lambda}+1}\left[
    K^{(\ell m)}+\frac{r-2M}{\lambda r+3M}\left\{ H_2^{(\ell m)}
    -r\partial_r K^{(\ell m)} \right\} \right]\nonumber\\
&&+\frac{(r-2M)}{\lambda r+3M}\left(r^2\partial_r
    G^{(\ell m)}-2h_1^{(\ell m)}\right)\ .
\end{eqnarray}

Price and Pullin\cite{PP94} have consistently
worked in the first perturbative order by
linearizing Misner data and evolving them with the Zerilli equation.
The excellent agreement with full numerical results up to values
of $\mu_0\approx1.8$ was somewhat unexpected.

To linearize the 3-metric (\ref{M3}) we can use the
perturbative notion of $\Phi^4-1\ll1$.
In this case, after expansion into Legendre polynomials, we get 
\begin{equation}\label{Fdef}
\Phi^4\approx
       1+8\left( 1+\frac{M}{2R}\right)^{-1}\!\!\!
       \sum_{\ell=2,4...}\!\!\!\kappa_\ell(\mu_0)
       \left(M/R\right)^{\ell+1}P_\ell(\cos\theta),
\end{equation}
where
\begin{equation}\label{kapdef}
       \kappa_\ell(\mu_0)\dot=
        \frac{1}{\left(4\Sigma_1\right)^{\ell+1}} \sum_{n=1}^\infty
        \frac{(\coth{n\mu_0})^\ell}{\sinh{n\mu_0}}
\end{equation}
Note that the expansion parameter, $M\coth(\mu_0)/(4R\Sigma_1)$,
has to be less than $1$ for the above expansion to be valid. This 
condition is always fulfilled outside the effective single hole
event horizon, located at $R=M/2$, for
$\mu_0<1.56865$. For bigger values of $\mu_0$ one has the inverse expansion
analogous to that made in Eq.\ (2.21) of Ref.\ \cite{LP97a}.

Since to first order we only have quadrupolar $(\ell=2)$ contributions,
expansion (\ref{Fdef}) has the nice feature of separating, in the common
factor $\kappa_2$, all the dependence on the initial distance between holes.
Thus allowing to compute the whole family of evolutions with only one
integration of the Zerilli equation (\ref{rtz}), and then rescale
waveforms by their corresponding $\kappa_2(\mu_0)$.
In the Regge--Wheeler notation one finds that the only
non-vanishing components of the perturbed 3-metric (\ref{M3})
are $H_2=K$. In this way one builds up
the initial value of $Q^\ell$ (with $\partial_tQ^\ell=0)$ and then evolve
to obtain waveforms, spectra, and radiated energies\cite{PP94}.

\begin{figure}
\epsfysize=2.8in \epsfbox{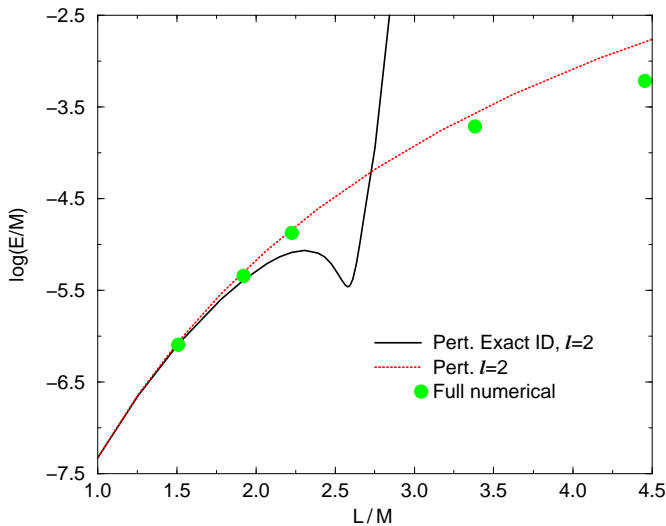}
\caption{The total gravitational radiation computed in the close limit
approximation via the Zerilli equation. The solid curve represents
the evolution of exact Misner initial data for two equal masses black
holes. At the bottom of the dip $(L\approx2.58)$ the system radiates one
order of magnitude less than that predicted by the evolution of linearized
initial data (dotted line). For comparison are given the results of full
numerical integration of Einstein equations.
}
\label{figZEL}
\end{figure}

It was first found by Abrahams and Price\cite{AP96a} that if one does not
linearizes the initial data, but extracts its multipoles from the exact
metric as
\begin{equation}\label{H2K}
H_2^{(\ell0)}=K^{(\ell0)}=\int d\theta\sin\theta\,\Phi^4Y_{\ell0},
\end{equation}
and evolves using the Zerilli equation, one obtains for the radiated energy
the disturbing results reproduced in Fig.\ \ref{figZEL}. While for small
separations, $L/M<2$, corresponding to $\mu_0<1.0$, the agreement with
full numerical computations\cite{AHSSS95}
and linearized perturbations is good; the energy curve
presents a local maximum at $\mu_0\approx1.25$ and then decreases to a local
minimum at $\mu_0\approx1.41$. At this point the system radiates an order of
magnitude less than the evolution of linearized initial data. For
larger values of $\mu_0$ the radiated energy rapidly increases and soon
after $\mu_0\approx1.49$, when it crosses upwards the linear prediction, it
overestimates the total energy radiated
by several orders of magnitude. All this happens well
before the linearized theory begins to deviate from the full numerical
results $(\mu_0>1.8)$. To unravel this paradoxical result we will first
compute the corresponding radiated energies in an alternative formulation
of the black hole perturbations.

\section{Curvature perturbations approach}

There is an independent formulation of the perturbation problem
derived from the Newman-Penrose formalism\cite{T73} that fully 
exploits the null structure of black holes allowing to
uncouple for a single wave equation to describe perturbations around
Kerr black holes. The outgoing gravitational radiation is fully 
described in this gauge (and tetrad) invariant formalism by the Weyl scalar
\begin{equation}
\psi_4=-C_{\alpha\beta\gamma\delta}\;n^\alpha\bar m^\beta
n^\gamma\bar m^\delta ,
\end{equation}
where $n^\mu$ and $\bar m^\mu$ (together with $l^\mu$ and $m^\mu$) form
the tetrad that span the spacetime.

$\psi_4$ fulfills the Teukolsky equation, which for the Schwarzschild case
reads
\begin{eqnarray}\label{master}
&&\ \ \Biggr\{\left(-\frac{r^4}\Delta \right)
\partial_{tt}+4r\left(1-\frac{Mr}\Delta \right) \partial_t  \nonumber \\
&&\ \ +\,\Delta^{2}\partial_r\left( \Delta^{-1}\partial_r\right) +
\frac 1{\sin \theta }\partial_\theta \left( \sin \theta \partial_\theta
\right) +\left( \frac 1{\sin^2\theta } \right) 
\partial_{\varphi \varphi }\nonumber\\
\ &&-\left[\frac{4i\cos \theta }{\sin^2\theta }
\right] \partial_\varphi -\left( 4\cot^2\theta +2\right) \Biggr\}r^4\psi_4
=0,  
\end{eqnarray}
where $\Delta=r^2-2Mr$.

The first step towards building up the initial $\psi_4$ and $\partial_t\psi_4$
is to find an instantaneous exact tetrad
compatible with the data (\ref{M3}). We have found it by
choosing the $l^\mu$ and $n^\mu$ that generate shear free null congruences
(spin coefficients $\sigma=0=\lambda)$ and fix the form of
$m^\mu$ and its complex conjugate $\bar m^\mu$
under transformations of type III (boosts) in such a way that the
spin coefficient $\epsilon=0$. Our tetrad is then
\begin{eqnarray}\label{tetrconf}
   (l^{\mu}) &=& \bigg(\frac{1}{1-2M/r},\Phi^{-2},0,0\bigg) \; ,\nonumber \\
   (n^{\mu}) &=& \frac{1}{2} \,\bigg(1,-(1-2M/r)\Phi^{-2},0,0\bigg) \; , \\
   (m^{\mu}) &=& \frac{\Phi^{-2}}{\sqrt{2}r}\,
                  \bigg(0,0,1,i/\sin\theta \bigg)\; .\nonumber 
\end{eqnarray}

Using formulae (3.1) and (3.2) of Ref.\ \cite{CLBKP98} that give
$\psi_4$ and $\partial_t\psi_4$ in terms of the 3-geometry and the extrinsic
curvature (here vanishing) we obtain\footnote{Note that Eqs.\ (3.1) and (3.2)
of Ref.\ \protect{\cite{CLBKP98}} are exact for our data
if we just drop the (0) and (1) labels. 
Note also an obvious misprint there: The second addend in Eq (3.1) and 
the third one in Eq (3.2) should carry a factor 8 instead of 4.}
\begin{eqnarray}
&&\psi_4\Big|_{t=0}=\frac{(1-2M/r)}{2r^2\Phi^6}
\left\{-3(\partial_\theta\Phi)^2+\Phi(\partial^2_\theta-\cot\theta\ 
\partial_\theta)\Phi\right\}\nonumber\\
\nonumber\\
&&\partial_t\psi_4\Big|_{t=0}=-\frac{2M}{r^2\Phi^2}\psi_4\Big|_{t=0}\; . 
\end{eqnarray}

\begin{figure}
\epsfysize=2.8in \epsfbox{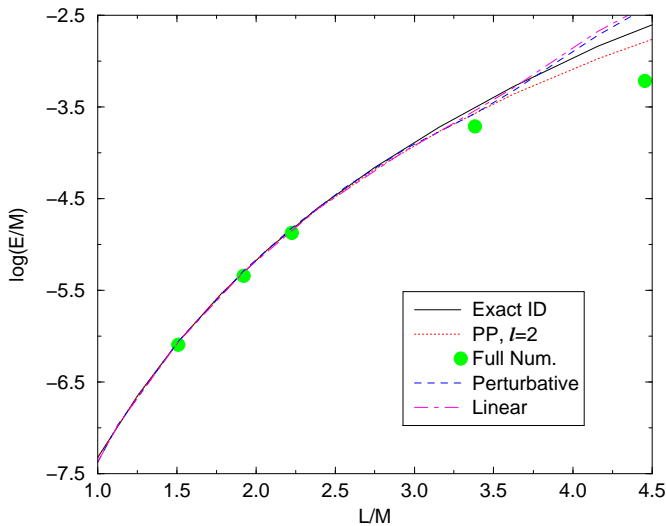}
\caption{The total radiated energy as computed using the
Newman-Penrose-Teukolsky approach. The solid line is the linear evolution of
the exact initial data for two black holes initially at rest. There is
complete agreement with full numerical results for $L<3.3$
and with the Price-Pullin results (labeled as PP, $\ell=2$) all the way up
to much bigger initial separations, keeping under control all the initial
nonlinearities. Also shown two other notions of perturbative initial data
in the Newman-Penrose formalism.
}
\label{TodoEL}
\end{figure}

The evolution of this initial data via the Teukolsky equation \cite{KLPA97}
produce
the results shown in Fig.\ \ref{TodoEL} for the total energy emitted
as gravitational waves. There is no dip at any value of the separation
of the holes. The predicted energy agrees with the full numerical results
for all $\mu_0<1.8$. For larger values of the separation 
(although we evolved here {\it exact} initial data for
two distinct black holes) we overestimate the
radiated energy at practically the same rate as the original Price--Pullin
result who extrapolated the linearized, $\ell=2$, initial data
to values of $(\mu_0>1.58)$ beyond the radius of convergence of their expansion
parameter, i.e. $M\coth(\mu_0)/(4R\Sigma_1)>1$. Correcting for this fact
produce the $\ell=2$ piece of the curve labeled as ``Linear''.

How robust is this result? To answer this question 
we fist considered different tetrad rotations of type III
to define the relative normalization of the vectors $l^\mu$ and
$n^\mu$ while keeping fixed its null directions.
The total radiated energy was quite insensitive to this choice. Some
choices slightly improved the agreement with full numerical results, but we
found not a priori justification for such choices. We thus keep tetrad
(\ref{tetrconf}),
which in the Schwarzschild limit reproduces the background tetrad\cite{T73}
used to write down the Teukolsky equation in Boyer--Lindquist coordinates.

We also essayed two possible definitions of perturbative notion in the
Newman--Penrose formalism. The first one, that we called {\it perturbative},
considers the exact Weyl tensor contracted with the background (Schwarzschild)
tetrad
\begin{eqnarray}
\psi_4^{\rm pert}\Big|_{t=0}
&=&\frac{(1-2M/r)(1+\Phi^4)}{4r^2\Phi^2}\nonumber\\
&&\times\left\{-3(\partial_\theta\Phi)^2+\Phi(\partial^2_\theta-\cot\theta\ 
\partial_\theta)\Phi\right\}\nonumber\\
\nonumber\\
\partial_t\psi_4^{\rm pert}\Big|_{t=0}
&=&-\frac{4M}{r^2(1+\Phi^4)}\psi_4\Big|_{t=0}\; , 
\end{eqnarray}
and a second possibility, that we called {\it linear} because it considers
only linear terms in the conformal factor $\Phi$, gives
\begin{eqnarray}
\psi_4^{\rm linear}\Big|_{t=0}&=&\frac{(1-2M/r)}{2r^2}
\left(\partial^2_\theta-\cot\theta\ \partial_\theta\right)\Phi\nonumber\\
\nonumber\\
\partial_t\psi_4^{\rm linear}\Big|_{t=0}
&=&-\frac{2M}{r^2}\psi_4\Big|_{t=0}\; . 
\end{eqnarray}
The resulting energy is essentially unchanged in the $\mu_0<1.8$ regime
and grows steeper than the exact initial data for larger initial separations
of the holes. This is because for the initial perturbative $\psi_4$'s
higher $\ell$ contributions are less under control than for the exact
initial $\psi_4$. This is seen in the waveforms extracted far away from
the system. While for the perturbative choices of $\psi_4$ waveforms look
fine up to $\mu\approx1.6$ those evolved from the exact $\psi_4$ reach
$\mu\approx2.0$ without much higher $\ell$ content.
No dip was
found at any value of the separation of the holes and in fact we could
{\it not} reproduce Fig.\ \ref{figZEL} results within the
Newman-Penrose-Teukolsky formalism. This gives us a measure
of a certain robustness of this approach
against nonlinearities included in the initial data.

\section{A closer look at the problem}

\begin{figure}
\epsfysize=2.8in \epsfbox{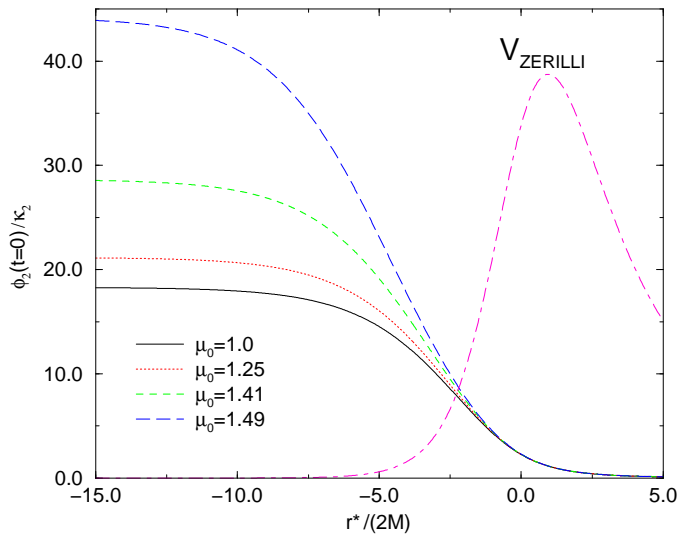}
\caption{The $\ell=2$-multipole projection of the exact Misner data is
used to build up the Moncrief-Zerilli initial waveform. Waveforms are
normalized by $\kappa_2$ in order to show only deviations from the linear
regime. Also shown is the Zerilli potential to stress that initial data
only differ well inside the potential barrier.
}
\label{figZID}
\end{figure}

Since both evolution equations, Zerilli's and Teukolsky's are equivalent
in the linear perturbations regime we are working in, the different results
obtained must be related to how $\psi_4$ and $Q$ handle the nonlinearities
included in the Misner data. We then first plot in  Fig.\ \ref{figZID} the
Moncrief waveform normalized to $\kappa_2$ such that in the linear regime
all curves would superpose. Still for $\mu_0=1.0$ the curve do
not deviate much from that of the linear data. As we increase the initial
separation of the holes we
find an steady increase of the amplitude in the region $r^*<0$ while for
$r^*>0$
the curves lie on top of each other. Most of the outgoing radiation is
generated around the top of the Zerilli potential $V_\ell$ at
$r^*/2M\approx0.95$.
The form of the initial data in terms of $Q_\ell$ do not show any particular
feature around $\mu_0=1.41$. Their relative form continues to grow in amplitude
very close to the horizon, but remains almost unchanged further outside.
We checked that the effect of the change of the expansion parameter
at $\mu_0=1.56865$ described above is {\it not} responsible for the dip
at $\mu_0=1.41$.

\begin{figure}
\epsfysize=2.8in \epsfbox{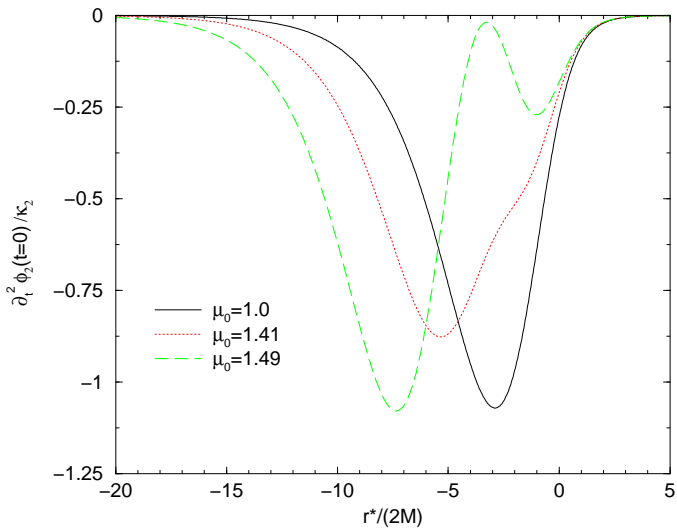}
\caption{The initial second time derivative of the Moncrief waveform normalized
by $\kappa_2$ to compare with the linear initial data. While there is a good
superposition of data form $r^*>0$, notable differences appear for $r^*<0$.
the maximum amplitude decreases steadily for $0<\mu_0<1.41$ and increases
for $\mu_0>1.41$. Also the location of the maximum amplitude monotonically
recedes towards more negative $r^*$ as $\mu_0$ increases. This provokes a
decrease of the frequency for which destructive interference of the
outgoing radiation occurs. 
}
\label{figZppID}
\end{figure}

\begin{figure}
\epsfysize=2.8in \epsfbox{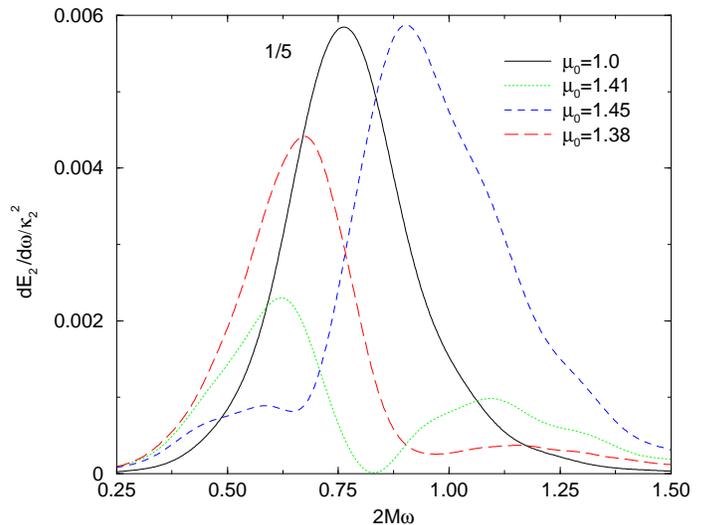}
\caption{The spectrum of the gravitational radiation far away from the system
as evolved by the Zerilli equation. The solid line labeled with the
separation parameter $\mu_0=1$ gives essentially the linear regime (we
rescaled its amplitude by one fifth).
we see how the suppression effect sets up sharply
around $\mu_0=1.41$. For  $\mu_0=1.38,\ 1.41,\ 1.45$ destructive interference
occurs around frequencies $2M\omega\approx0.95,\ 0.83,\ 0.65$ respectively.
}
\label{figZSP}
\end{figure}

The quantity directly related to radiation $\partial_tQ$ initially vanishes for
Misner data (time symmetric), in Fig.\ \ref{figZppID} we plot thus
$\partial_t^2Q$ at $t=0$. Observe again the good superposition for $r^*>0$.
There is also a big depression for linear data with a minimum at
$r^*/2M\approx-3$. As the two black holes start
more separated and $\mu_0$ increases, the amplitude of this depression decreases
and its location recedes towards more negative $r^*$'s. This continues so
until we reach $\mu_0\approx1.41$. Then the amplitude begins to
grow while still the location of the depression recedes
towards negative $r^*$'s.
Is this relatively small decrease of the amplitude (around 20\%) responsible
for one order of magnitude less radiated energy?
Actually since the depression is located very close to the horizon, well inside
the potential barrier, most of the radiation generated is swallowed by the
black hole and just a very small piece of it reaches infinity.
To answer the above question one has to take into account the wave nature
of the gravitational radiation. We know that a great deal of the radiation
coming out to infinity is generated around the maximum of the Zerilli
potential at $r^*_{max}/2M\approx0.95$. There will also be a piece of the
disturbance generated at around the pick of $\partial_t^2Q$. these two pulses
will be out of phase by $(2M\omega)(\Delta t/2M)$ where $\Delta t$ is the
time the pulse generated close to the horizon takes to arrive at
$r^*/2M\approx0.95$. Depending on their relative phases these two pulses
can produce constructive or destructive interference. Assuming 
$\Delta t\sim\Delta r^*$, where $\Delta r^*$ is the distance between the
pick of $\partial_t^2Q$ and the maximum of the Zerilli potential.
In the linear regime, this analysis show that the destructive interference
appears for frequencies $2M\omega>1.2$; too high to influence the
total energy radiated at infinity. As we increase $\mu_0$, the pick in
Fig.\ \ref{figZppID} moves to the left, thus generating destructive
interference at lower frequencies. In Fig.\ \ref{figZSP} we show that this
effect is clearly visible in the three spectra for $\mu_0=1.38,\ 1.41,\ 1.45$.
There is destructive interference around
$2M\omega\approx0.95,\ 0.83,\ 0.65$ respectively.
The strong suppression of the radiation at $\mu_0=1.41$ is then due to
destructive interference right at the frequency of the maximum of the linear
spectrum, at $2M\omega\sim0.8$.

By the same mechanism we can explain the sudden increase in the radiated
energy for $\mu_0>1.4$. It is now the effect of destructive interference
acting at too low frequencies and constructive interference at higher ones
together with a dramatic increase of the amplitude of the initial data with
respect to the linear regime. Finally for $\mu_0>1.56865$ the effect of
the change in the perturbative parameter makes things blow up.

Alternatively one can look at the initial form of $\psi_4$ and its time
derivative computed via the linear transformations given in
Refs.\ \cite{CL98,CKL98} in terms of the initial Moncrief waveform $\psi$.
This transformation can be seen as part of the process of 
linearization that takes place during the evolution with the Zerilli
equation. We observe then in Fig.\ \ref{figZppID} the same receding
of the picks as the initial separation of the holes increase, and
its separation from the maximum of the Zerilli potential agrees very
well with the prediction coming from the interference patterns seen
in Fig.\ \ref{figZSP}.
We will see that this behavior is in contrast with that of
the direct computation of the exact $\psi_4$ and $\partial_t\psi_4$.

\begin{figure}
\epsfysize=2.8in \epsfbox{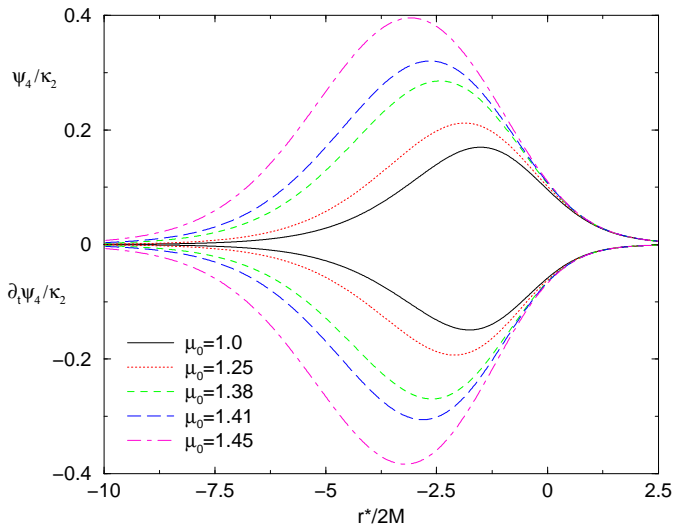}
\caption{Initial $\psi_4$ and $\partial_t\psi_4$ derived from the
Moncrief waveform $\phi_2$ via Eq.\ (2.9) of Ref.\ \protect\cite{CKL98}
for different initial separations of the holes.
}
\label{IDMonP4}
\end{figure}

\section{Discussion}

\begin{figure}
\epsfysize=2.7in \epsfbox{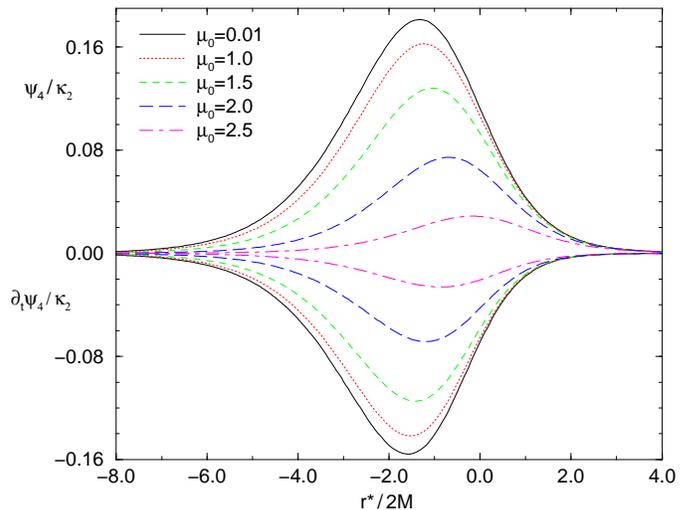}
\caption{The exact initial $\psi_4$ and $\partial_t\psi_4$ for the Misner
geometry. The effect of increasing $\mu_0$ is to decrease the amplitude and
shift the maximum towards less negative $r^*$. Curves tend to superpose in
the region more relevant for radiation, $r^*>0$.
}
\label{figIDP4}
\end{figure}

Unraveled the mechanism that generates the Abrahams --Price dip, there remains
the question of why nothing like this happens when we use the
Newman--Penrose--Teukolsky approach to Schwarzschild perturbations. Again,
a plot of the initial data for different separations leads to the answer.
In Fig.\ \ref{figIDP4} we observe that the maximum of the initial data
for increasing $\mu_0$ shifts toward {\it increasing} $r^*$'s instead of
more negative ones as happened for the Zerilli formalism. Thus regardless of
a small decrease in the amplitude the destructive interference occurs at
too high frequencies to influence the outgoing radiation.

The study of exact initial data evolved linearly has taught us several
lessons: i) How the initial Misner data vary with respect to their linearized
version and which aspects of them are relevant for computation of the
gravitational radiation that reaches infinity. ii) The interference
effects that occurs when one evolve these initial data. Note that the
occurrence of interference is in general independent of nonlinearities. In fact
it is rather sensitive to the shift of the maximum that they produce in
the particular case of Misner data. iii) Since the Zerilli--Moncrief waveform
and $\psi_4$ are two intrinsically different objects, they respond differently
to nonlinearities. We have seen that for the initial data studied in this
paper $\psi_4$ is a better suited quantity to evolve linearly.

There is still another way of evolving Misner data. namely nonlinearly,
by direct full numerical integration of Einstein equations. In this way we
of course evolve nonlinear initial data and obtain the correct behavior.
In Ref.\ \cite{BBCLST99}
we used the code {\it Cactus} to evolve the full set of Einstein
equations for a single distorted black hole. We have found that {\it even}
for small linear initial distortions we needed to solve the initial value
problem nonlinearly, otherwise errors in satisfying the constraints
would generate numerically unstable evolutions.

Our results are relevant to the idea of evolving fully numerically
binary black holes starting from large separations (starting from
Post-Newtonian initial data)
until a common horizon encompasses the system and then let
perturbation theory to take over\cite{LP00}.
The perturbation taking over full numerical
method has the advantage of optimizing
supercomputer resources, concentrating them in the region where
the two black holes are completely
detached and full nonlinear relativistic effects take
place. Once a common horizon forms one can assume the close limit
approximation to hold and continue the evolution with
a single wave equation on the background of a Kerr black hole.
It is very fortunate that is the Newman-Penrose-Teukolsky approach
rather than the Regge-Wheeler-Zerilli one that has this nice behavior in
response to
nonlinear Cauchy data since the Teukolsky equation can be generalized
to rotating black hole backgrounds while the metric
perturbation approach do not.
Unless there is a dramatic progress in full numerical technics, this
perturbative--full numerical hybrid,
is our only chance to get waveform templates for inspiraling black holes
before laser interferometers begin to operate.
The effect described in Fig.\ \ref{figZEL} inhibit us from
using the Zerilli equation to that end. On the other hand,
the Teukolsky evolution seems better suited for this marriage between
full numerical and perturbative approaches. We also checked that
the same general bad behavior of the Zerilli equation and healthy one
for the Teukolsky equation holds when one considers Brill-Lindquist
initial data (other solution of the conformally flat,
time symmetric initial value problem). If this behavior is also true
in the astrophysically more appealing scenario of orbiting black holes
is currently under investigation. If we remain within the Bowen--York
family of initial data (conformally flat and Longitudinal)
we expect a decomposition of the conformal factor of the
type\cite{Nicasio:1998aj} $\Phi=\Phi_{Misner}+\Phi_{reg}$, where
$\Phi_{reg}$ is proportional to the square of the momentum of the holes
and the square of their distance. This means that at least for black
holes with small initial momentum the effects discussed
in this paper should qualitatively still be present.

Another situation where our results should be considered
is when one is interested in studying
second order perturbations of rotating black holes\cite{CL99}. The
perturbative approach to the Newman-Penrose formalism form a hierarchy
of equations
\begin{equation}
\widehat{{\cal T}}\psi_4^{(N)}={\cal S}[\psi
^{(N-1)},\partial _t\psi ^{(N-1)}].  \label{dos}
\end{equation}

where ${\cal T}$ stands for the background wave operator of the
Teukolsky equation, $\psi_4^{(N)}$ is the waveform of the $N$ 
perturbative order considered,
and ${\cal S}$ is a source term formed by products of all 
perturbations of order lower than $N$.

One can solve the above equations by evolving initial data order-by-order,
successively reaching the next perturbative stage. An alternative approach
to that is to evolve exact initial data with the first order
wave equation (it has a vanishing source term in vacuum), and then
evolve second
order equations with vanishing initial data (expressing the source in
terms of the first order perturbations). This is consistent to the
desired second perturbative order.
The justification to this approach can be found from a Laplace transform
analysis. In the appendix of paper\ \cite{CL98} it is explicitly
found that the initial data dependence can be included in 
the Teukolsky equation as an additional source term, ${\cal S}_{ID}$ and this
holds at each order $N$ in the above hierarchy of equations (\ref{dos}).
Since the additional source term adds linearly
and the wave operator is always the background one,
summing over all $N$ orders corresponds to evolve the first order equation
with the {\it exact} initial data. Then one evolves all higher
orders with {\it vanishing} initial data.

\begin{acknowledgments}
The author thanks M. Campanelli for discussions that lead me to undertake
the studies developed here.
C.O.L. is a member of the Carrera del Investigador Cient\'\i fico 
of CONICET, Argentina.
\end{acknowledgments}

\end{document}